# Quantum oscillation in carrier transport in two-dimensional junctions

*Junfeng Zhang[1,2], Weiyu Xie[2], Michael L. Agiorgousis[2], Duk-Hyun Choe[2], Vincent Meunier[2], Xiaohong, Xu[1], Jijun Zhao[3*], and Shengbai, Zhang[2*]*

[1] Research Institute of Materials Science of Shanxi Normal University & Collaborative Innovation Center for Shanxi Advanced Permanent Magnetic Materials and Technology, Linfen 041004, China

[2] Department of Physics, Applied Physics, and Astronomy, Rensselaer Polytechnic Institute, Troy, NY 12180, USA

[3] Key Laboratory of Materials Modification by Laser, Ion and Electron Beams (Dalian University of Technology), Ministry of Education, Dalian 116024, China

**KEYWORDS**. Two-dimensional structures, Vertical and lateral junctions; Quantum transport; Quantum Oscillation; Quantum well potential.

**ABSTRACT**. Two-dimensional (2D) device structures have recently attracted considerable attention. Here, we show that most 2D device structures, regardless vertical or lateral, act as a lateral monolayer-bilayer-monolayer junction in their operation. In particular, a vertical structure cannot function as a vertical junction as having been widely believed in the literature. Moreover, due to a larger electrostatic screening, the bilayer region in the junction always has a smaller band gap than its monolayer counterpart. As a result, a potential well, aside from the usual potential barrier, will form universally in the bilayer region to affect the hole or electron quantum transport in the form of transmission or reflection. Taking black phosphorus as an example, we show that an oscillation in the transmission coefficient can be clearly resolved in a two-electrode prototypical device by non-equilibrium Green function combined with density functional theory calculations and the results can be qualitatively understood using a simple quantum well model. The presence of the quantum well is vital to 2D device design, including the effective tuning of quantum transmission by a vertical electric field.

Two-dimensional (2D) materials are known to have unique properties, which may display great potentials for novel applications. One important area is in the electronic device where significant effort has been made to integrate different 2D materials into vertical or lateral junctions, as a basic component of the nanodevices such as field-effect transistor (FET), *p-n* junction, and photovoltaics. The vertical and lateral junctions using 2D materials such as graphene, *h*-BN, black phosphorous (BP), and transition metal dichalcogenides (TMDs), have been fabricated in laboratories [1-7].

However, in practical applications, most of 2D junctions are neither pure vertical nor pure lateral as often assumed in theory[8, 9], but a monolayer (ML)-bilayer (BL)-ML (ML-BL-ML) structure. In experiment, it is difficult to make an electrode contact to only one ML without metallization between the electrode and the entire 2D stack. Even if such a contact can be made, one cannot avoid current tunneling between the layers[10-17]. Hence, most realistic 2D vertical junctions follow the schematic plot shown in Figure 1a, where the overall structure can be viewed as a three-junction device with two lateral and one vertical junctions in series. It is therefore not surprising that an intended vertical *p-n* junction may not behave as a vertical diode but rather as lateral one[14, 16]. For 2D lateral junctions fabricated by the two-step chemical vapor deposition (CVD) method[18-23], it is also common to obtain structures with partial overlap of the MLs to form a common BL region at the interface, especially when the component materials have a large lattice mismatch[24-26]. In some of the recent van der Waals 2D junctions[27-29], graphene has been used as part of the electrodes for lower contact resistance. This, however, does not change the fact that the device is always a multi-junction device, as we will discuss below.

We will define the ratio of the BL region over the ML regions (which for simplicity is assumed to be symmetric, see Figure 1a) as $R_{B/M} = L_{BL}/L_{ML}$, in which $L_{BL}$ and $L_{ML}$ are the lengths of the BL and ML regions, respectively. With such a definition, a vertical junction may be classified as a special case of the lateral junctions when $R_{B/M} \to \infty$. Even in this case, the two lateral junctions to the electrodes cannot be ignored. It is thus clear that the fundamental device physics of 2D junctions such as the band offsets and associated transport properties can deviate significantly from those predicted for either pure vertical or pure lateral structures.

Figures 1a-b show two prototypical 2D junctions with either two components[25, 26, 30, 31] (heterojunction between A and B, i.e., A-AB-B) or a single component[32, 33] (homojunction between A and A, i.e., A-AA-A). An important property of 2D semiconducting materials is that a BL or few-layer structure always has a smaller band gap than the ML. Because of this, a potential well naturally forms in the BL region for either electrons or holes, e.g., in Figures 1c-d. In conventional 3D superlattices, the effects of quantum well (QW) have been extensively studied [34, 35] In contrast, its effects on the carrier transport in 2D junctions have attracted little attention.

In this paper, by a combined first-principles non-equilibrium Green function (NEGF) and density functional theory (DFT) calculation of model 2D junction devices, we show a universal oscillation in the low-energy carrier transmission modes due to the presence of the QW. The calculations are done using elemental black phosphorus (BP) as an example for it is an elemental semiconductor with a high carrier mobility (up to ~1,000 cm$^2$ V$^{-1}$ s$^{-1}$)[36, 37] and a large band gap dependent on layer thickness[37, 38]. The oscillation can be understood by using a one-dimensional QW model, which yields results in agreement with DFT calculations. Both theories suggest that the length of the QW, which may vary from atomic scale to the mean free path of electrons (e.g.,

~10 nm in graphene and TMD materials[39, 40]), can be critical to the conductance of the device. We further show that a gate placed on top of a 2D conducting channel, as in a standard FET setting, is sufficient to create the potential well to control carrier transport.

Supercells of 2D BP junctions were constructed by joining ML and AB-stacked BL nanoribbons at zigzag interfaces, as shown in Figure 2a. We used a large supercell length of 8.27 nm (18 unit cells in lateral) to ensure that a reasonably good bulk band structure can be reached in the centers of both the ML and BL regions. The atomic structures and electronic properties were calculated using DFT and the projector-augmented wave (PAW) method[41], as implemented in the Vienna *ab initio* simulation package (VASP)[42, 43]. The Perdew–Burke–Ernzerhof (PBE)[44] functional was used to describe the exchange-correlation interactions. The cutoff energy for the plane-wave expansion was set to 400 eV. The Brillouin zones were sampled by **k**-point grids with a uniform separation of 0.015 Å$^{-1}$. A 3 nm thick vacuum region in the direction perpendicular to the 2D BP plane was added to avoid spurious interactions between periodic images. The dangling bonds at the BL edges were passivated by hydrogen atoms. The atoms were fully relaxed until the maximum forces on the atoms were less than 0.01 eV/Å.

To study transport properties, we constructed model devices with two electrodes, either placed in the same ML or in different MLs, as shown in Figures S1a-b. We adopted Au electrodes on both sides for their good conductivity. The scattering region was fixed at ~ 9 nm and the length of the BL varied from 1.1 to 4.5 nm. The transmission coefficients and local density of states (LDOS) were computed using the Keldysh-NEGF formalism[45, 46] combined with first-principles method, as implemented in the Nanodcal code. Real space linear combination of DZP atomic orbital basis was employed. The quantum transmission were calculated by including self-energies for the coupling of scattering region to the semi-infinite Au leads under the zero-bias voltage. The cutoff energy for the real space grid was set to 400 eV. A $1 \times 21 \times 1$ **k** point mesh was employed in the Brillouin zone integration for transport calculations.

We start from DFT calculations for the A-AA-A BP homojunction, while the particular choice of systems should not affect the generality of our findings. When the lengths of the devices are sufficient, the behavior of the 2D junctions should obey the Anderson limit[8]. Hence, here we focus on the junction with a shorter length, i.e., a supercell of 10 nm length for a 2D homojuction, made of an AB-stacked BL and a ML BP (see Figure 2a for the structural model). From DFT calculations of the combined ML-BL BP system, the band gaps in the bulk ML and BL regions are 0.97 and 0.69 eV (see Figure 2b), respectively, which are compared to the values of 0.88 and 0.59 eV for bulk ML and BL BPs[47, 48]. In accordance, the valence and conductance band offsets of the 2D homojuction are 0.39 and 0.11 eV, as displayed in Figure 2c. Here we focus on the former (corresponding to hole transport), as a larger band offset could make it easier to observe the QW effect.

One can understand the QW effect on transport by standard quantum mechanics for a single particle where the potential $v(x)$ is $-V_0$ inside the well and zero outside. Using the boundary conditions, the transmission coefficient ($T$) of a carrier can be readily derived as[49]:

$$T = [1 + \frac{V_0^2}{4E(E+V_0)} \cdot \sin^2(\frac{L_{BL}}{\hbar}\sqrt{2m(E+V_0)})]^{-1} \qquad (1)$$

where $L_{BL}$ is the length of the potential well, $m$ is the carrier mass, and $E$ is the energy of carrier. With a fixed $V_0$, the calculated $T$ is shown in Figure 3a (lower panel), showing a *sinusoidal* oscillation with respect to $L_{BL}$. At a low energy, e.g., $E < 50$ meV, $T$ oscillates quickly between 0 to 1. However, when the energy is higher, $T$ approaches a constant of 1, irrespective of $L_{BL}$. Most strikingly, low-energy carriers could be totally blocked ($T = 0$, OFF state) or transmitted ($T = 1$, ON state) depending on the choice of $L_{BL}$.

To assess the prediction by such a simple quantum mechanics model, we perform a series of NEGF calculations on two-electrode devices made of 2D BP homojunctions. Two examples are considered here: the stacked (Figure S1a) and staggered (S1b) junctions, where the edges have been passivated by hydrogen atoms (the effect of different passivation will be discussed below). In either case, a potential well is formed in the BL region for holes. The difference between the two is that the transmitted electrons in the staggered junction must participate in a vertical interlayer transport. The transport properties are obtained with a fixed scattering region length but a varying $L_{BL}$, as illustrated in Figure S1a. Figure S2 shows, as an example for $L_{BL} \sim 3$ nm, the LDOS in the scattering region in stacked and staggered 2D BP junctions. Metallic behavior can be readily seen in both $x < 1$ nm and $> 8$ nm regions of the BPs, suggesting a buffer-region effect from the metal electrodes. Consistent with the DFT results in Figure 2b, the valence band maximum (VBM) of the BL region acts as a potential well (which is labeled with dashed lines in Figure S2). It is noteworthy that the shape of the potential well deviates from that of the square-well model due to the relaxation of charge at interface.

Figure 4 shows transmission coefficients for both homojunctions with $L_{BL}$ = 1–4.5 nm. In our calculation, the system is periodic along the interface (i.e., in the $y$ direction), which results in a band dispersion. Hence, the depth of the potential well also varies with $k$ points as shown in Figure 2b. For example, the well depth is 0.39 eV at $\Gamma$, but only 0.07 eV at $M$. Consequently, carriers with different momenta would experience wells at different depths. For simplicity, we consider $\Gamma$ point scattering only, as holes are mostly localized here. An oscillating behavior can be clearly seen in Figure 4 for both stacked and staggered junctions. For comparison, results of the QW model in Eq. (1) are also given, which qualitatively agree with NEGF results, showing also an oscillation with $L_{BL}$. For a more quantitative comparison, let us define standard deviations, $\delta_1$ and $\delta_2$, where $\delta_1$ is the deviation of the DFT results with respect to the QW-model results and $\delta_2$ is the deviation of the DFT results from the average value. We find that $\delta_1$ is noticeably smaller than $\delta_2$, i.e., $\delta_1/\delta_2 = 37.7\%$ for the stacked case (Figure 4a) and 44% for the staggered case (Figure 4b), respectively. Although some deviations from the QW model due to its oversimplification can be expected, it captures the essential physics well, as supported by the DFT-NEGF calculations.

Next, we study the effect of edges: to this end, we (1) replaced all hydrogen by fluorine and (2) removed all hydrogen leaving behind bare edges. We found that the potential well is very much unchanged, except near its edges. Importantly, in Figure S3, we still see oscillations in the transmission coefficient, similar to those in Figure 4. Hence, we conclude that edge passivation is nonessential, as it at best only modestly alters the details of quantum oscillations in the transmission coefficient.

Note that the existence of a potential well is not limited to BP or a homojunction, but should be more general for 2D junctions. For example, for a heterojunction made of two materials (A-AB-B) (see Figure 1b), the symmetric potential well will be replaced by an asymmetric one, as shown in Figure 3b. The QW model predicts that (see details in SI)

$$T = \frac{16 k_1 k_2^2 k_3}{k_2^2 (k_1+k_3)^2 + (k_2^2 - k_1^2)(k_2^2 - k_3^2) \sin^2 k_2 a} \times \frac{k_2 k_3}{(k_2 + k_3)^2} \quad , \tag{2}$$

in which $k_1 = \sqrt{2m(E+V_1)}/\hbar$, $k_2 = \sqrt{2m(E+V_2)}/\hbar$, $k_3 = \sqrt{2mE}/\hbar$, where $m$, $\hbar$, $L_{BL}$, and $E$ are already defined in Eq (1), $V_1$ is the band offset between material A and B, and $(V_2 - V_1)$ is the potential well indicated in Figure 3b. If A = B, $V_1$ vanishes and $(V_2 - V_1)$ becomes $V_0$ in Eq. (1). Similar to the homo case, the oscillatory behavior of $T$ is also evident in Figure 3b (the lower panel with $V_1$ = 0.05 eV and $V_2$ = 0.15 eV). Different from the homo case, however, the maximum transmission coefficient here can hardly approach 1 when $E$ is small, as also indicated in Figure S4. As long as the BL region has a smaller band gap than the ML region, an oscillation in $T$ exists.

Note that the above oscillation in the transmission coefficient should not be limited to 2D junctions. In a homogeneous bilayer or few-layer 2D material, if one modulates the band gap, e.g., by applying an external electric field[50, 51] or a strain[52, 53], a similar behavior can also be expected. In a recent experiment, it has been demonstrated that the band gap of few-layer BP can be tuned from 300 meV down to 50 meV by ramping up the (effective) electric field from 0 to 1.1 V/nm[51]. While our DFT calculation on an AA-stacked BL BP shows a band gap reduction from 500 meV to 100 meV when the electric field reaches 3.5 V/nm. Figure S5 shows that, at the maximum field, potential well depths of 0.18 eV at the conductance band minimum (CBM) and 0.2 eV at the VBM are created. Based on this, we propose a prototypical homogeneous BL quantum-well device, in which the well depth can be controlled by gating, as demonstrated in Figure 5. In general, quantum oscillation may occur in a 2D device provided that the band gap of the 2D material is spatially in-homogeneously tuned. The exceptions are that if the transport is non-ballistic (i.e., when the scattering region is larger than the mean free path of the carriers) or the energy of the carriers is too high so that the effect of the quantum well diminishes. This controllable quantum-well effect provides a different route in designing ultra-high ON/OFF ratio transport devices.

To summarize, our analysis suggests that most of the 2D junctions experimentally realizable today are neither a purely vertical junction nor a purely lateral junction, but take the hybrid ML-BL-ML form. Our study using DFT-NEGF, combined with analytical quantum mechanics model, unveils the unexpected roles of the BL region. Rather than functioning as a vertical junction as often believed, the BL acts as a quantum well to result in quantum oscillations in the transmission. Both the magnitude and period of the oscillations are strong functions of the width of the BL region so they should be readily detectable in experiment. Furthermore, our study points to the general importance of controlling the spatial homogeneity in 2D devices for desired transport properties.

## ASSOCIATED CONTENT

**Supporting Information**.

The stacked and staggered structure models used in the NEGF calculations (Figure S1). The LDOS maps of the stacked and staggered structures with $L_{BL}$ ~ 3 nm (Figure S2). The transmission coefficients as a function of $L_{BL}$ (Figure S3). The maximum and minimum transmission coefficients from the symmetric and asymmetric QW models, respectively (Figure S4). The CBM and VBM of the AA-stacked BL BP, as a function of the applied effective electric field (Figure S5). Details on the quantum-well model (last section).

This material is available free of charge via the Internet at http://pubs.acs.org.

## AUTHOR INFORMATION

**Corresponding Author**

*Email: zhaojj@dlut.edu.cn (Jijun Zhao), zhangs9@rpi.edu (S. B. Zhang).## ACKNOWLEDGMENT

<>JFZ was supported by the National Natural Science Foundation of China (11304191), the Natural Science Foundation of Shanxi province (2015021011) and CSC Award No. 201608140024. JJZ was supported by the Fundamental Research Funds for the Central Universities of China (DUT16LAB01), WYX was supported by the US-NSF under Award No. 1104786 and SBZ was supported by the US-DOE under Grant No. DESC0002623. The supercomputer time by NERSC under DOE contract No. DE-AC02-05CH11231 and by the CCI at RPI are also acknowledged.

REFERENCES

1. Geim, A. K.; Grigorieva, I. V. *Nature* **2013,** 499, (7459), 419-425.

2. Lim, H.; Yoon, S. I.; Kim, G.; Jang, A.-R.; Shin, H. S. *Chem. Mater.* **2014,** 26, 4891-4903.

3. Wang, H.; Liu, F.; Fu, W.; Fang, Z.; Zhou, W.; Liu, Z. *Nanoscale* **2014,** 6, (21), 12250-12272.

4. Liu, Y.; Weiss, N. O.; Duan, X.; Cheng, H.-C.; Huang, Y.; Duan, X. *Nat. Rev. Mater.* **2016,** 1, 16042.

5. Pant, A.; Mutlu, Z.; Wickramaratne, D.; Cai, H.; Lake, R. K.; Ozkan, C.; Tongay, S. *Nanoscale* **2016,** 8, (7), 3870-3887.

6. Li, D.; Chen, M.; Sun, Z.; Yu, P.; Liu, Z.; Ajayan, P. M.; Zhang, Z. *Nat. Nanotechnol.* **2017,** 12, 901-906.

7. Zhao, J.; Cheng, K.; Han, N.; Zhang, J. *WIREs Comput. Mol. Sci.* **2017**, e1353.

8. Zhang, J.; Xie, W.; Zhao, J.; Zhang, S. *2D Mater.* **2017,** 4, 015038.

9. Yu, H.; Kutana, A.; Yakobson, B. I. *Nano Lett.* **2016,** 16, (8), 5032-5036.

10. Roy, T.; Tosun, M.; Cao, X.; Fang, H.; Lien, D.-H.; Zhao, P.; Chen, Y.-Z.; Chueh, Y.-L.; Guo, J.; Javey, A. *ACS Nano* **2015,** 9, (2), 2071-2079.

11. Chuang, S.; Kapadia, R.; Fang, H.; Chang, T. C.; Yen, W.-C.; Chueh, Y.-L.; Javey, A. *Appl. Phys. Lett.* **2013,** 102, (24), 242101.

12. Furchi, M. M.; Pospischil, A.; Libisch, F.; Burgdörfer, J.; Mueller, T. *Nano Lett.* **2014,** 14, (8), 4785-4791.

13. Deng, Y.; Luo, Z.; Conrad, N. J.; Liu, H.; Gong, Y.; Najmaei, S.; Ajayan, P. M.; Lou, J.; Xu, X.; Ye, P. D. *ACS Nano* **2014,** 8, (8), 8292-8299.

14. Cheng, R.; Li, D.; Zhou, H.; Wang, C.; Yin, A.; Jiang, S.; Liu, Y.; Chen, Y.; Huang, Y.; Duan, X. *Nano Lett.* **2014,** 14, (10), 5590-5597.

15. Wang, F.; Wang, Z.; Xu, K.; Wang, F.; Wang, Q.; Huang, Y.; Yin, L.; He, J. *Nano Lett.* **2015,** 15, (11), 7558-7566.

16. Lee, C.-H.; Lee, G.-H.; van der Zande, A. M.; Chen, W.; Li, Y.; Han, M.; Cui, X.; Arefe, G.; Nuckolls, C.; Heinz, T. F.; Guo, J.; Hone, J.; Kim, P. *Nat. Nanotechnol.* **2014,** 9, (9), 676-681.

17. Doan, M.-H.; Jin, Y.; Adhikari, S.; Lee, S.; Zhao, J.; Lim, S. C.; Lee, Y. H. *ACS Nano* **2017,** 11, (4), 3832-3840.

18. Yu, Q.; Jauregui, L. A.; Wu, W.; Colby, R.; Tian, J.; Su, Z.; Cao, H.; Liu, Z.; Pandey, D.; Wei, D.; Chung, T. F.; Peng, P.; Guisinger, N. P.; Stach, E. A.; Bao, J.; Pei, S.-S.; Chen, Y. P. *Nat. Mater.* **2011,** 10, (6), 443-449.

19. Liu, L.; Park, J.; Siegel, D. A.; McCarty, K. F.; Clark, K. W.; Deng, W.; Basile, L.; Idrobo, J. C.; Li, A.-P.; Gu, G. *Science* **2014,** 343, (6167), 163-167.

20. Duan, X.; Wang, C.; Shaw, J. C.; Cheng, R.; Chen, Y.; Li, H.; Wu, X.; Tang, Y.; Zhang, Q.; Pan, A.; Jiang, J.; Yu, R.; Huang, Y.; Duan, X. *Nat. Nanotechnol.* **2014,** 9, (12), 1024-1030.

21. Huang, C.; Wu, S.; Sanchez, A. M.; Peters, J. J. P.; Beanland, R.; Ross, J. S.; Rivera, P.; Yao, W.; Cobden, D. H.; Xu, X. *Nat. Mater.* **2014,** 13, (12), 1096-1101.

22. Gong, Y.; Lin, J.; Wang, X.; Shi, G.; Lei, S.; Lin, Z.; Zou, X.; Ye, G.; Vajtai, R.; Yakobson, B. I.; Terrones, H.; Terrones, M.; Tay, Beng K.; Lou, J.; Pantelides, S. T.; Liu, Z.; Zhou, W.; Ajayan, P. M. *Nat. Mater.* **2014,** 13, (12), 1135-1142.

23. Duan, X.; Wang, C.; Shaw, J. C.; Cheng, R.; Chen, Y.; Li, H.; Wu, X.; Tang, Y.; Zhang, Q.; Pan, A.; Jiang, J.; Yu, R.; Huang, Y.; Duan, X. *Nat. Nanotechnol.* **2014,** 9, (12), 1024-1030.

24. Gong, Y.; Lei, S.; Ye, G.; Li, B.; He, Y.; Keyshar, K.; Zhang, X.; Wang, Q.; Lou, J.; Liu, Z.; Vajtai, R.; Zhou, W.; Ajayan, P. M. *Nano Lett.* **2015,** 15, (9), 6135-6141.

25. Guimarães, M. H. D.; Gao, H.; Han, Y.; Kang, K.; Xie, S.; Kim, C.-J.; Muller, D. A.; Ralph, D. C.; Park, J. *ACS Nano* **2016,** 10, (6), 6392-6399.

26. Ling, X.; Lin, Y.; Ma, Q.; Wang, Z.; Song, Y.; Yu, L.; Huang, S.; Fang, W.; Zhang, X.; Hsu, A. L.; Bie, Y.; Lee, Y.-H.; Zhu, Y.; Wu, L.; Li, J.; Jarillo-Herrero, P.; Dresselhaus, M.; Palacios, T.; Kong, J. *Adv. Mater* **2016,** 28, (12), 2322-2329.

27. Avsar, A.; Vera-Marun, I. J.; Tan, J. Y.; Watanabe, K.; Taniguchi, T.; Castro Neto, A. H.; Özyilmaz, B. *ACS Nano* **2015,** 9, (4), 4138-4145.

28. Lee, G.-H.; Cui, X.; Kim, Y. D.; Arefe, G.; Zhang, X.; Lee, C.-H.; Ye, F.; Watanabe, K.; Taniguchi, T.; Kim, P.; Hone, J. *ACS Nano* **2015,** 9, (7), 7019-7026.

29. Roy, T.; Tosun, M.; Kang, J. S.; Sachid, A. B.; Desai, S. B.; Hettick, M.; Hu, C. C.; Javey, A. *ACS Nano* **2014,** 8, (6), 6259-6264.

30. Zhou, R.; Ostwal, V.; Appenzeller, J. *Nano Lett.* **2017**.

31. Liu, X.; Wei, Z.; Balla, I.; Mannix, A. J.; Guisinger, N. P.; Luijten, E.; Hersam, M. C. *Science Advances* **2017,** 3, (2).

32. Zhang, C.; Chen, Y.; Huang, J.-K.; Wu, X.; Li, L.-J.; Yao, W.; Tersoff, J.; Shih, C.-K. *Nat. Commun.* **2016,** 7, 10349.

33. Lin, H.-F.; Liu, L.-M.; Zhao, J. *J Mater Chem C* **2017,** 5, (9), 2291-2300.

34. Hicks, L. D.; Dresselhaus, M. S. *Phys. Rev. B* **1993,** 47, (19), 12727-12731.

35. Kuo, Y.-H.; Lee, Y. K.; Ge, Y.; Ren, S.; Roth, J. E.; Kamins, T. I.; Miller, D. A. B.; Harris, J. S. *Nature* **2005,** 437, (7063), 1334-1336.

36. Li, L.; Yu, Y.; Ye, G. J.; Ge, Q.; Ou, X.; Wu, H.; Feng, D.; Chen, X. H.; Zhang, Y. *Nat. Nanotechnol.* **2014,** 9, (5), 372-377.

37. Qiao, J.; Kong, X.; Hu, Z.-X.; Yang, F.; Ji, W. *Nat. Commun.* **2014,** 5, 4475.

38. Liu, H.; Neal, A. T.; Zhu, Z.; Luo, Z.; Xu, X.; Tománek, D.; Ye, P. D. *ACS Nano* **2014,** 8, (4), 4033-4041.

39. Cui, Q.; Ceballos, F.; Kumar, N.; Zhao, H. *ACS Nano* **2014,** 8, (3), 2970-2976.

40. Lu, C.-P.; Li, G.; Watanabe, K.; Taniguchi, T.; Andrei, E. Y. *Phys. Rev. Lett.* **2014,** 113, (15), 156804.

41. Kresse, G.; Joubert, D. *Phys. Rev. B* **1999,** 59, (3), 1758-1775.

42. Kresse, G.; Furthmüller, J. *Comp. Mater. Sci.* **1996,** 6, (1), 15-50.

43. Kresse, G.; Furthmüller, J. *Phys. Rev. B* **1996,** 54, (16), 11169-11186.

44. Perdew, J. P.; Burke, K.; Ernzerhof, M. *Phys. Rev. Lett.* **1996,** 77, (18), 3865-3868.

45. Taylor, J.; Guo, H.; Wang, J. *Phys. Rev. B* **2001,** 63, (12), 121104.

46. Wang, J.; Guo, H. *Phys. Rev. B* **2009,** 79, (4), 045119.

47. Guan, J.; Zhu, Z.; Tománek, D. *Phys. Rev. Lett.* **2014,** 113, (4), 046804.

48. Lei, S.; Wang, H.; Huang, L.; Sun, Y.-Y.; Zhang, S. *Nano Lett.* **2016,** 16, (2), 1317-1322.

49. Griffiths, D. J., *Introduction to quantum mechanics*. Second ed.; Pearson Educaion: New Jersey, 2005; p 78-82.

50. Liu, Q.; Zhang, X.; Abdalla, L. B.; Fazzio, A.; Zunger, A. *Nano Lett.* **2015,** 15, (2), 1222-1228.

51. Deng, B.; Tran, V.; Xie, Y.; Jiang, H.; Li, C.; Guo, Q.; Wang, X.; Tian, H.; Koester, S. J.; Wang, H.; Cha, J. J.; Xia, Q.; Yang, L.; Xia, F. *Nat. Commun.* **2017,** 8, 14474.

52. Rodin, A. S.; Carvalho, A.; Castro Neto, A. H. *Phys. Rev. Lett.* **2014,** 112, (17), 176801.

53. Fei, R.; Yang, L. *Nano Lett.* **2014,** 14, (5), 2884-2889.

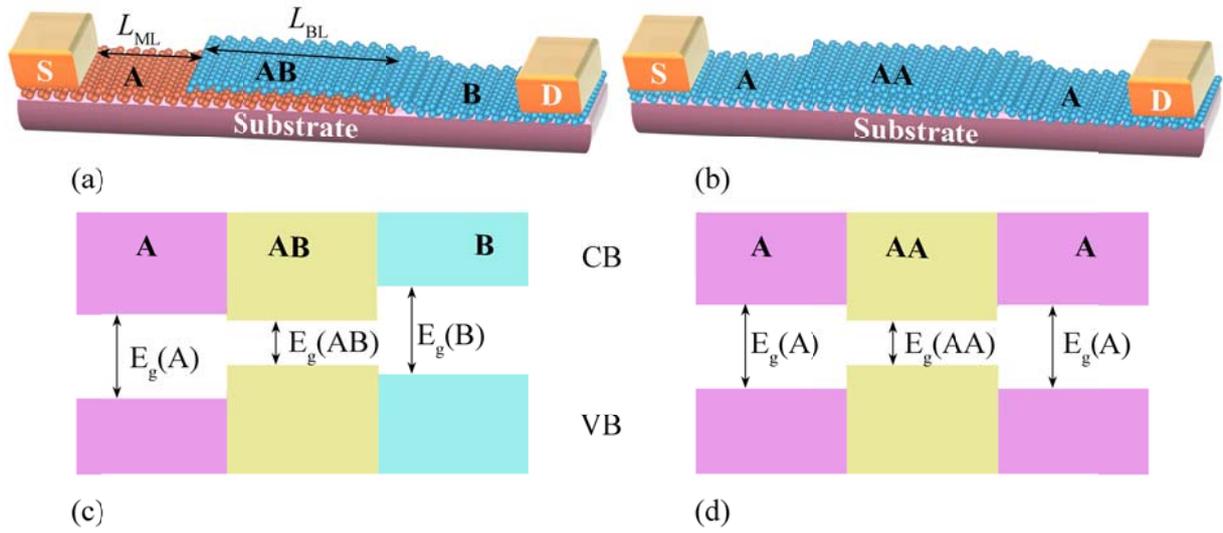

**Figure 1.** Typical structures of two-electrode devices with (a) an asymmetric (A-AB-B) and (b) a symmetric (A-AA-A) 2D junction. S and D stand for source and drain electrodes. (c) and (d) Their corresponding schematic band alignments. CB and VB stand for conduction and valence band, respectively.

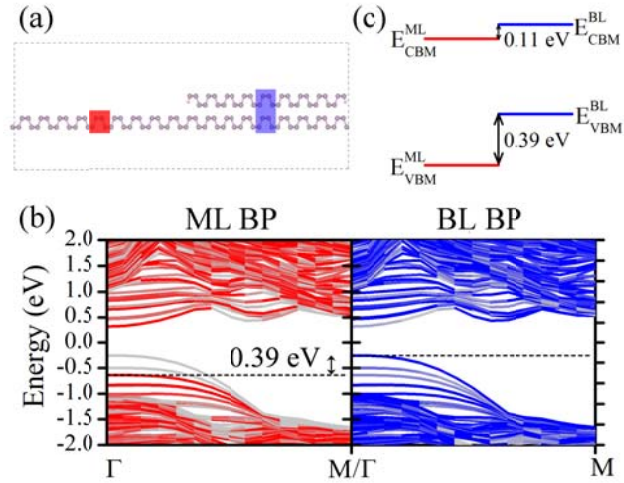

**Figure 2.** (a) The structural model of 2D BP homojunction used in the DFT calculation. Filled red and blue squares denote the central unit cells in the ML and BL domains, respectively. (b) Band structure of a 2D BP homojunction with zigzag edges. We performed atomic projection for them onto the (red and blue) central unit cells in (a): states with 10% projection in the ML red square are shown in red, BL states with 10% projection in the BL blue square are shown in blue, and those with less than 10% projections are shown in grey. (c) Band alignments in 2D BP homojunction determined from the band structures in (b). The 0.39 eV valence band offset between the ML and BL regions is calculated using the PBE functional.

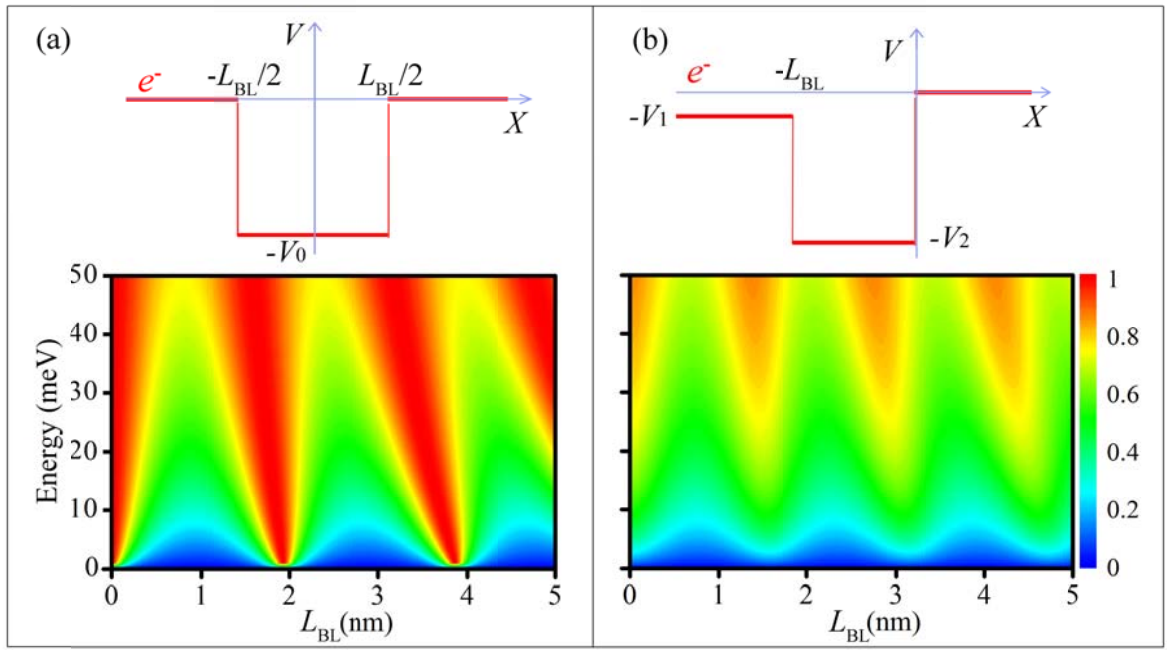

**Figure 3.** Schematic plots of the quantum well models (upper panel) and the corresponding transmission coefficients (lower panel) for (a) symmetric (A/AA/A) and (b) asymmetric (A/AB/B) 2D homojunctions. In (a), the well depth $V_0$ is set at 0.1 eV. In (b), however, the depths are $V_1 = 0.05$ and $V_2 = 0.15$ eV.

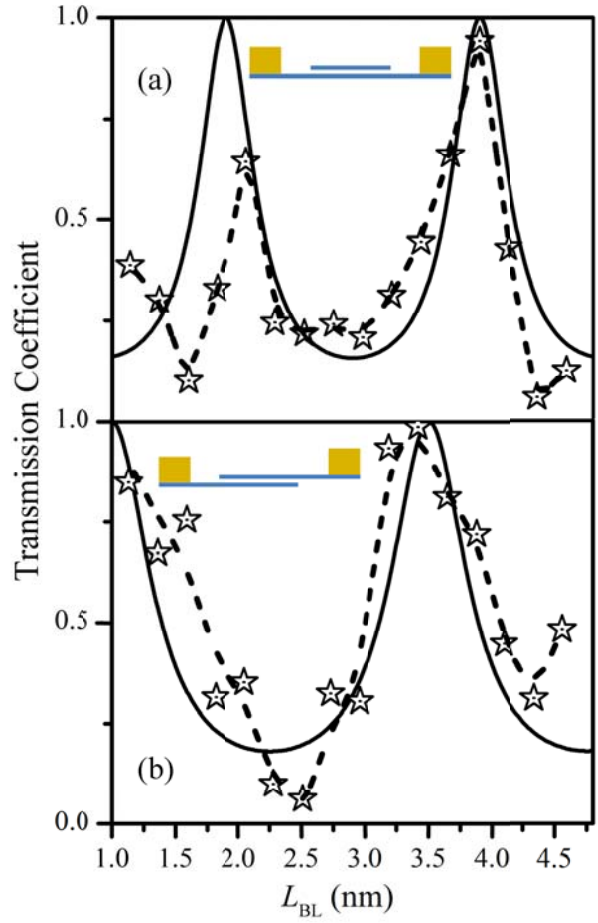

**Figure 4.** Transmission coefficients of the quantum-well model (black lines) and the DFT-NEGF calculations (stars) for (a) stacked and (b) staggered BP junctions. Insets show the schametic models for transport calculations. Dashed lines linking the stars are drawn to guide eyes. In the calculation, kinetic energy of the carriers and the well depth were set to 80 and 4 meV in (a), and 57 and 3 meV in (b), respectively.

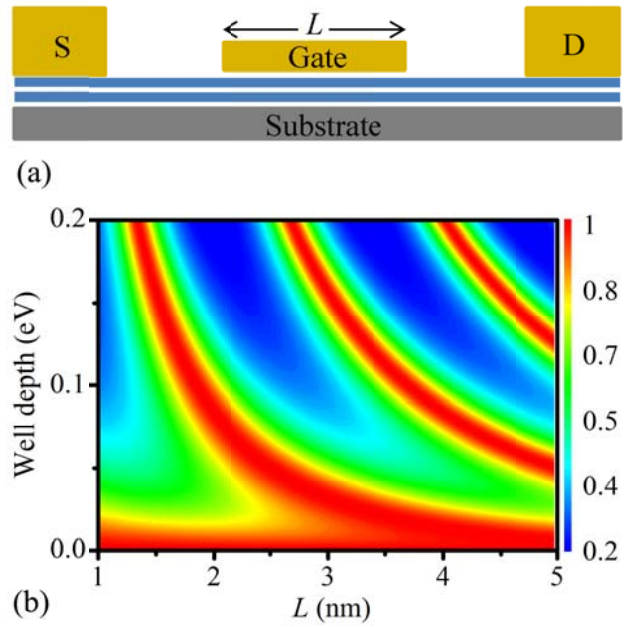

**Figure 5.** (a) A prototypical homogeneous BL quantum-well device. (b) Transmission coefficients (see Eq. 1) as functions of the well length ($L$) and depth (which is set by the gate voltage). For clarity, the kinetic energy of the carriers is set at 10 meV.

# Supporting Information

Quantum oscillation in carrier transport in two-dimensional junctions

*Junfeng Zhang[1,2], Weiyu Xie[2], Michael L. Agiorgousis[2], Duk-Hyun Choe[2], Vincent Meunier[2], Xiaohong, Xu[1], Jijun Zhao[3*], and Shengbai, Zhang[2*]*

[1]Research Institute of Materials Science of Shanxi Normal University & Collaborative Innovation Center for Shanxi Advanced Permanent Magnetic Materials and Technology, Linfen 041004, China

[2]Department of Physics, Applied Physics, and Astronomy, Rensselaer Polytechnic Institute, Troy, NY 12180, USA

[3]Key Laboratory of Materials Modification by Laser, Ion and Electron Beams (Dalian University of Technology), Ministry of Education, Dalian 116024, China

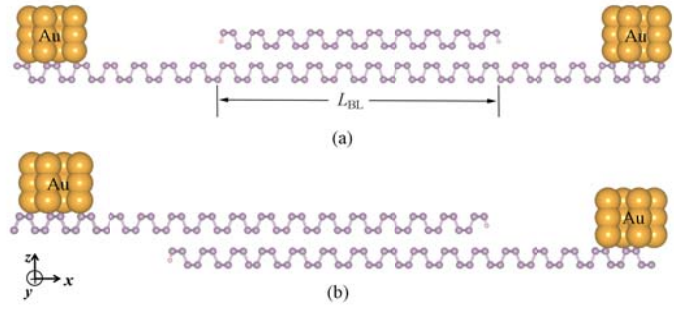

**Figure S1.** Structure models of (a) stacked and (b) staggered 2D BP homojunction used in the NEGF calculations. $L_{BL}$ stands for the length of the bilayer region.

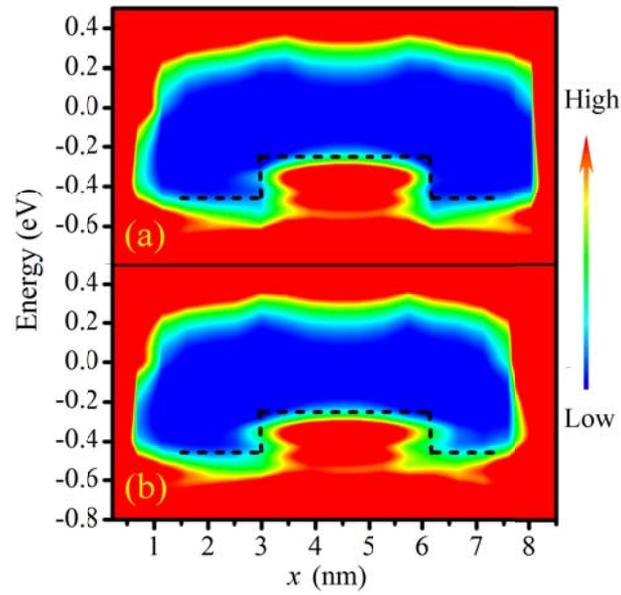

**Figure S2.** LDOS maps for stacked (a) and staggered (b) homojunction from DFT-NEGF calculations. The dashed line shows bilayer region (~ 3 nm) and the potential well induced by the valence band offset. The density of states increase from the blue to red as labeled in the right side.

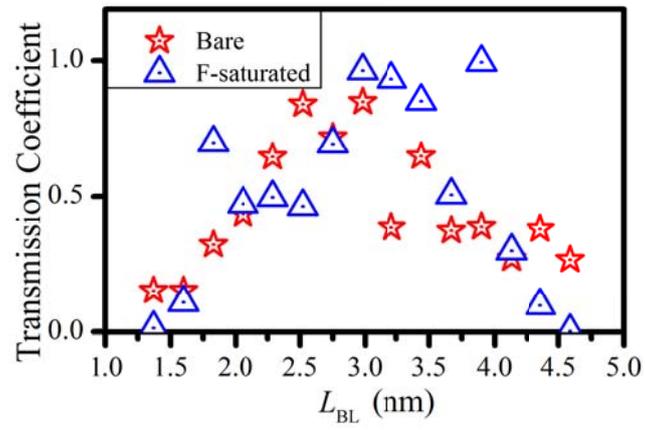

**Figure S3.** Transmission coefficients of stacked 2D BP heterojunctions with bare (red star) and fluorine passivated (blue triangle) edges.

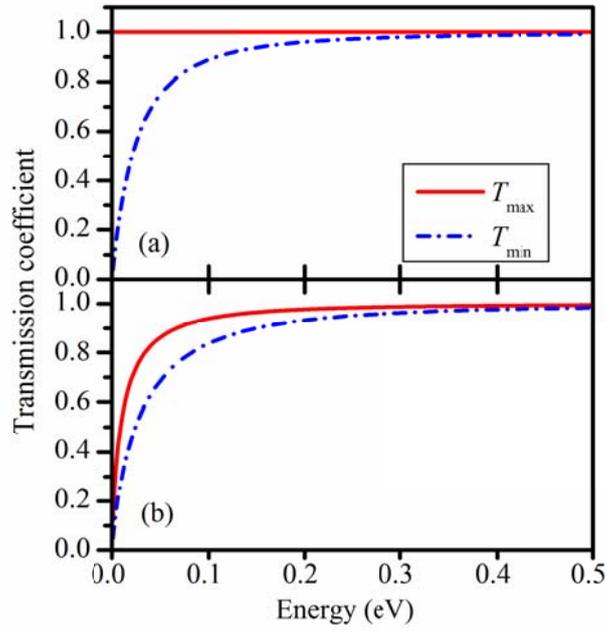

**Figure S4.** The maximum and minimum transmission coefficients from (a) symmetric and (b) asymmetric QW models, respectively.

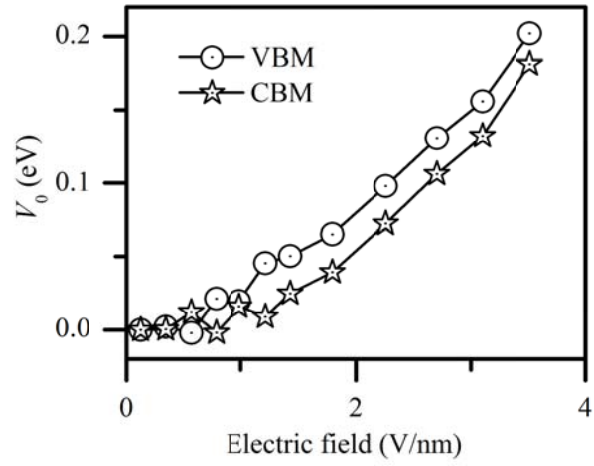

**Figure S5.** The VBM and CBM potential well depth ($V_0$) of AA stacked BL BP with the function of the external effective electric field.

**Details of the quantum well model.**

We first consider the symmetric potential well. For the finite square well as shown in Figure 4a,

$$v(x) = \begin{cases} 0, & \text{for } x < -L_{BL}/2 \\ -V_0, & \text{for } -L_{BL}/2 < x < L_{BL}/2 \\ 0, & \text{for } x > L_{BL}/2 \end{cases} \quad (S1)$$

where $V_0$ is the potential well depth, and $L_{BL}$ is the bilayer region length (potential well length). Here, we only consider the scattering states (with $E > 0$). We can write the Schrödinger equation says: $-\frac{\hbar^2}{2m}\frac{d^2\Psi}{dx^2} + v(x)\Psi = E\Psi \rightarrow \frac{d^2\Psi}{dx^2} = k^2\Psi$, where $k = \frac{\sqrt{2m(E-v(x))}}{\hbar}$. The general solution gives:

$$x < -L_{BL}/2 : \psi_1(x) = Ae^{ikx} + Be^{-ikx}, \; k = \frac{\sqrt{2mE}}{\hbar}$$
$$-L_{BL}/2 < x < L_{BL}/2 : \psi_2(x) = C\sin(lx) + D\cos(lx), \; k_2 = \frac{\sqrt{2m(E+V)}}{\hbar} \quad (S2)$$
$$x > L_{BL}/2 : \psi_3(x) = Fe^{ikx}, \; k = \frac{\sqrt{2mE}}{\hbar}$$

Then, the transmission coefficient is obtained by considered the boundary conditions:

$$T = |F|^2/|A|^2 = [1 + \frac{V_0^2}{4E(E+V_0)} \cdot \sin^2(\frac{L_{BL}}{\hbar}\sqrt{2m(E+V_0)})]^{-1} \quad (S3)$$

Next, we consider the asymmetric case as shown in Figure 4b. The potential can be expressed as

$$v(x) = \begin{cases} -V_1, & \text{for } x < -L_{BL} \\ -V_2, & \text{for } -L_{BL} < x < 0 \\ 0, & \text{for } x > 0 \end{cases} \quad (S4)$$

Similar with that in symmetric one, we can obtain the Schrodinger equations and get the general solutions as follows:

$$x < -L_{BL}: \quad \psi_1(x) = Ae^{ik_1 x} + Be^{-ik_1 x}, \; k_1 = \frac{\sqrt{2m(E+V_1)}}{\hbar}$$

$$-L_{BL} < x < 0: \psi_2(x) = Ce^{ik_2 x} + De^{-ik_2 x}, \; k_2 = \frac{\sqrt{2m(E+V_2)}}{\hbar} \quad . \quad (S5)$$

$$x > 0: \quad \psi_3(x) = Fe^{ik_3 x}, \quad k_3 = \frac{\sqrt{2mE}}{\hbar}$$

The total transmission coefficient can be obtained then:

$$T = \frac{16 k_1 k_2^2 k_3}{k_2^2 (k_1+k_3)^2 + (k_2^2 - k_1^2)(k_2^2 - k_3^2)\sin^2 k_2 a} \times \frac{k_2 k_3}{(k_2+k_3)^2} \quad . \quad (S6)$$